\documentclass{article}

\usepackage{arxiv}

\usepackage[utf8]{inputenc} \usepackage[T1]{fontenc}    \usepackage{hyperref}       \usepackage{url}            \usepackage{booktabs}       \usepackage{amsfonts}       \usepackage{nicefrac}       \usepackage{microtype}      \usepackage{lipsum}
\usepackage{graphicx}
\graphicspath{ {./combined/} }

\usepackage{graphicx}
\usepackage[utf8]{inputenc}
\usepackage{caption}
\usepackage{amssymb}
\usepackage{amsmath}
\usepackage{algorithm}
\usepackage[noend]{algorithmic}
\usepackage{cite}
\usepackage{float}
\usepackage{caption,subcaption}

\usepackage{siunitx}
\usepackage{booktabs}

\def\citeyear{\cite} \def\endnote{\footnote}

\title{Open Data and Quantitative Techniques for Anthropology of Road Traffic}

\author{
Ajda Pretnar Žagar \\
  Faculty of Computer and Information Science\\
  University of Ljubljana\\
  Večna pot 113, 1000 Ljubljana, Slovenia\\
  \texttt{ajda.pretnar@fri.uni-lj.si} \\
\And
Tomaž Hočevar \\
  Faculty of Computer and Information Science\\
  University of Ljubljana\\
  Večna pot 113, 1000 Ljubljana, Slovenia\\
  \texttt{tomaz.hocevar@fri.uni-lj.si} \\
\And
Tomaž Curk \\
Faculty of Computer and Information Science\\
  University of Ljubljana\\
  Večna pot 113, 1000 Ljubljana, Slovenia\\
  \texttt{tomaz.curk@fri.uni-lj.si} \\
}

\date{}

\begin{document}
\maketitle

\begin{abstract}
What kind of questions about human mobility can computational analysis help answer? How to translate the findings into anthropology? We analyzed a publicly available data set of road traffic counters in Slovenia to answer these questions. The data reveals interesting information on how a nation drives, how it travels for tourism, which locations it prefers, what it does during the week and the weekend, and how its habits change during the year. We conducted the empirical analysis in two parts. First, we defined interesting traffic spots and designed computational methods to find them in a large data set. As shown in the paper, traffic counters hint at potential causes and effects in driving practices that we can interpret anthropologically. Second, we used clustering to find groups of similar traffic counters as described by their daily profiles. Clustering revealed the main features of road traffic in Slovenia. Using the two quantitative approaches, we outline the general properties of road traffic in the country and identify and explain interesting outliers. We show that quantitative data analysis only partially answers anthropological questions, but it can be a valuable tool for preliminary research. We conclude that open data are a useful component in an anthropological analysis and that quantitative discovery of small local events can help us pinpoint future fieldwork sites. \end{abstract}


\section*{Introduction}

With computational techniques gaining popularity in anthropology, there is no lack of examples successfully mixing quantitative and qualitative approaches. Ethno-mining~\cite{anderson2009}, stitching~\cite{blok2014complementary}, blending~\cite{bornakke2018bigthick}, circular mixed methods~\cite{pretnar2019data}, or hybrid methodologies~\cite{cury2019hybrid} join the best of both (methodological) worlds into a single research framework for exploring human habits and practices. However, could one do predominantly quantitative anthropological research?

Anthropology has been, at least since Malinowski~\cite{sredniawa1981anthropologist}, deeply intertwined with long periods spent with the people under research, \emph{i.e.}, fieldwork, and with a particular research technique called participant observation~\cite{spradley2016participant}. The researcher participates in the community's daily life with a heightened awareness of social, cultural, economic, religious, and physical processes. Such participation results not only in detail-rich narratives called ethnographies but also in a lived, embodied experience~\cite{csordas1990embodiment}. The researcher in anthropology is often seen as a ``research tool'' - a medium that translates observations into a coherent and structured form.

Therefore, can one do anthropological research without immediate human contact but solely using semi-big data and computational techniques? To establish if such research is possible and how to conduct it, we analyzed a publicly available data set recording traffic frequencies on roads in Slovenia. We aimed to determine national driving patterns and outline Slovenian road traffic. We argue that there is valuable information in traffic data showing how a nation drives, travels for tourism, which locations it prefers, what it does during the week and the weekend, and how the nation's habits change during the year. We corroborate these findings with newspaper archive data on local festivities, road closures, and infrastructure changes.

First, we established the typical traffic behaviour, i.e., we looked at how a nation usually moves around the country. Exploratory analysis showed when traffic peaks occur and where, what are the most frequented locations, and how the traffic changes during the weekend and between different seasons. Second, we used statistical methods to find interesting counters, namely those locations that showed a deviation from the typical behaviour (\emph{e.g.}, road accidents, traffic congestions). Unusual traffic patterns show how traffic changes and hint at potential causes and effects that we can interpret anthropologically. Third, we use clustering to structure the traffic patterns into groups and again try to explain those groups from the perspective of mobility and traffic as a reflection of social structures. This analysis is an example of how publicly available data, combined with computational techniques, could be used by anthropologists to gauge human practices. To encourage reproducibility and reuse of the data, we provide Jupyter notebooks and Orange workflows\endnote{\href{https://figshare.com/s/5a74f668444fedf51b4f}{Figshare repository with workflows and scripts}}. In conclusion, we discuss if and how quantitative analysis could be used in anthropology.

\section*{Related work}

Traffic is a reflection of human habits and practices~\cite{kuipers2013rise,podjedbezjak2017}. It is a form of communication where the locations (cities, villages, points of interest) are the emitters and receivers, and the drivers are the signal. As Horta~\citeyear{horta2019peripatetic} succinctly argues, roads embody a social process in which the collective life of the society emerges to the surface through movement. In our case, traffic movement is a social process with specific characteristics, intent, and implications for society.

Estevan~\citeyear{estevan1994contra} argues that ``mobility'' is a quantitative side of the movements performed by people and commodities, while ``accessibility'' is the qualitative counterpart. Thus, traffic data describe ``mobility,'' the basic patterns of social movement, quantity, frequency, and (i)regularity. However, glimpses of ``accessibility'' are also hidden in quantitative data. The activity of individuals is a reflection of social structures - collective preferences, negotiations, restrictions, and affordances. Traffic data is an aggregated individual action, displaying social structures through preferred (most frequented) or shunned (least frequented) locations, handling of (traffic) contingencies and corresponding strategies, and broad habitual patterns of the nation.

For example, as Yazıcı~\citeyear{yazici2013towards} argues, highways hold particular social significance, as they reflect social transformations in terms of economic shifts and infrastructural changes. Highways provide connections to major cities, enable daily migration (commuting), and empower the periphery. By looking at the expansion, deprecation, and modulation of highways and regional roads towards larger cities, one can infer how people commute and how economic opportunities change in time.

Cars are, in a way, an extension of human beings, affording them specific mobility and motility~\cite[ability of spontaneous movement]{dant2004driver}. Assemblages of drivers and cars do not embody the technology, but the relationship between society and technology and how technology is used and appropriated to achieve specific means. These specific means, \emph{i.e.}, the intentions behind driving are demonstrated in the choice of destination, route, and driving time are also contained in a numerical form of traffic data.

Several studies demonstrate the usefulness of quantitative analysis of human mobility data. Palmer et al.~\citeyear{palmer2013new} relate GPS positioning and demographic data to establish intimate patterns of human mobility, showing activity, segregation, and well-being of research participants through spatial data. Focusing on urban areas, Gallotti et al.~\citeyear{gallotti2021unraveling} analyze mobility data from 10 large cities to determine the structure of internal flows. Their analysis uncovers the heterogeneity of flows and internal segregation of cities, which aids in determining future policy responses. With London as their use case, Aslam and Cheng~\citeyear{aslam2018smart} model individual points of interest using the data from the Oyster Smart Card (payment system for the London underground). Taking it a step further, Zhao et al.~\citeyear{zhao2021characteristics} use high-resolution human mobility data from cell phones. They show that high-resolution data differ significantly from more high-level aggregated mobility data.

Nevertheless, when high-resolution data is not available, even low-resolution data can reveal interesting patterns. Huang and Wong~\citeyear{huang2015modeling} extracted regular individual activity from the geotagged Twitter data, showing that low-resolution data can provide valuable insights. Traffic data from the present study fall into the medium resolution category and, as we show, offer comparable insights into human mobility, for example, the difference in mobility between weekdays and weekends or dependence of the time of the day on location~\cite{palmer2013new}. 

In anthropology, quantitative data for analyses of human mobility have been used sparingly. Podjed~\citeyear{podjed2017augmented} shows how to combine ethnography with quantitative data analysis to determine the differences in national driving styles and the effect of mobile apps on driving behaviour. The author used quantitative data to map frequent behaviours, while qualitative data from the interviews and the so-called ``participant driving'' was used to explain them. In the same volume, Babič~\citeyear{babic2017discovering} presents an ethnographic analysis of traffic as reflected in the media discourse. Linguistic analysis can be qualitative or quantitative, and while Babič's analysis was qualitative and manual, it could also be computational with text mining and natural language processing. A combination of the two approaches would be the third option, likely generating additional insight into the discourse on traffic. Mixed methods work well for various research problems~\cite{blok2014complementary,pretnar2019data}, but no research has yet attempted a predominantly quantitative anthropological study. In the paper, we present an experimental quantitative anthropological analysis and explore how to look at traffic data as a cultural and social phenomenon. Our focus is not on devising new analytical methods but on discovering novel insights in existing data sets. We argue that even a medium resolution open data set can serve as a helpful starting point for anthropological research.

\section*{Open data in anthropology}

In anthropology, most of the data consists of interviews, field notes, and lived experiences, some of which cannot be translated to tabular data. Some contain personal and sensitive information. Thus, it is no wonder why anthropological open data repositories are rare, but they are not nonexistent.

In 1949, five American universities established the Human Relations Area Files (HRAF), a non-profit organization providing access to major databases on cultural diversity. eHRAF, its main database, is an index of worldwide cultures and human societies based on the identifier codes from the Outline of Cultural Materials (OCM). While eHRAF, the organization's main product with online access to the index, is accessible only with membership privileges, the site also hosts an open-source database of cross-cultural studies, Explaining Human Culture\endnote{\href{https://hraf.yale.edu/ehc}{Explaining Human Culture}}. Similarly, AnthroBase\endnote{\href{http://www.anthrobase.com/}{AnthroBase}} is a searchable database of anthropological texts, but it is a result of community effort, not institutional engagement.

Outside of anthropology, there is an abundance of services offering open access data\endnote{\href{https://libguides.umn.edu/c.php?g=858128}{An overview of OA repositories by the University of Minnesota}}, for example, OpenDOAR\endnote{\href{http://v2.sherpa.ac.uk/opendoar/}{OpenDOAR}} and Registry of Open Access Repositories (ROAR\endnote{\href{http://roar.eprints.org/}{Registry of Open Access Repositories}}). Both provide a list of public databases with statistics on the content and upload activity, making it easy to shortlist relevant services. CESSDA\endnote{\href{https://datacatalogue.cessda.eu/}{CESSDA}} is a database for social sciences, with open access to survey and questionnaire results, demographic data, experiment results, and more. Results of qualitative and multi-method studies are available at the Qualitative Data Repository (QDR\endnote{\href{https://qdr.syr.edu/}{Qualitative Data Repository}}), whose aim is also to prevent the data in social sciences from going to waste.

Most modern governments nowadays offer at least some type of access to public information. In continuation, we present an experiment where we took publicly available data, quantitatively analyzed them, and tried to interpret the findings from an anthropological point of view. We present some main results and expose the strengths and shortcomings of quantitative analyses in anthropology.

\section*{Data}

Analysis of traffic flows uses the traffic data from the Slovenian Infrastructure Agency~\cite{drsi}. The agency coordinates an extensive network of road traffic counters, which are installed on motorways and regional roads~\cite[on ways to measure traffic]{bickel2007measuring}. The data is publicly available through an application programming interface (API) providing real-time data and directly from the agency, which stores historical data. We analyzed the historical data for 2015, 2016, and 2017, for which traffic counts are available at hourly intervals. The traffic is reported per type of vehicle. We decided to look at cars and motorbikes, excluding buses, lorries, and trucks, as we were interested in the mobility patterns of the population and not in freight or public transport. Additionally, traffic counters report traffic for each direction in the case of regional roads and each lane in the case of motorways, making it necessary to consider the direction of each measurement.

There are 903 traffic counters, each reporting for two directions or lanes. We removed the counters with a significant proportion of missing data, i.e., the counters that did not report data for all twelve months or had a fall-out (all values for a particular month were zero). We ended up with 654 counters with sufficient longitudinal data. For the greater part of the analysis, we aggregated the data so that each counter reports average traffic frequencies (per month, season, or day of the week). We removed the data in the 10\% and 90\% quantile when averaging, thus excluding extraneous factors that could affect the results. Where we distinguish between weekdays and weekends, we explicitly state it. Weekend days were Saturdays, Sundays, and public holidays. We make this separation because the traffic patterns are significantly different in these two periods. With all aggregations done, this resulted in several hundred thousand rows of monthly data. To analyze the data by hand, we would have to inspect numerous combinations of counter IDs, days of the week, and directions to find counters with interesting patterns. To avoid manually browsing interesting counters and arbitrarily deciding what is interesting, we used three statistical techniques to measure the ``interestingness'' of the line plots showing monthly averages. 

Sample scripts and workflows are available on figshare\endnote{\href{https://figshare.com/s/5a74f668444fedf51b4f}{Figshare repository with workflows and scripts}}. Workflows can be loaded in Orange~\cite{JMLR:demsar13a}, a software for machine learning and data mining. They replicate the visualizations that are presented here. 

\section*{Methods}

For the needs of this research, we defined the interestingness of the counter as an unusual monthly, daily (by day of the week), or hourly deviation from the general traffic trend. Our primary assumption is that traffic is a reasonably predictable phenomenon, which shows distinctive patterns throughout the year. In general, traffic counters show consistent trends, mainly corresponding to commuter traffic. However, some of them have hourly spikes, seasonal increases, or monthly deviations that are interesting. They identify shifts in expected traffic with implications for the immediate locality and sometimes even the entire country.

\begin{figure}[ht]
\centering
\includegraphics[width=0.96\linewidth]{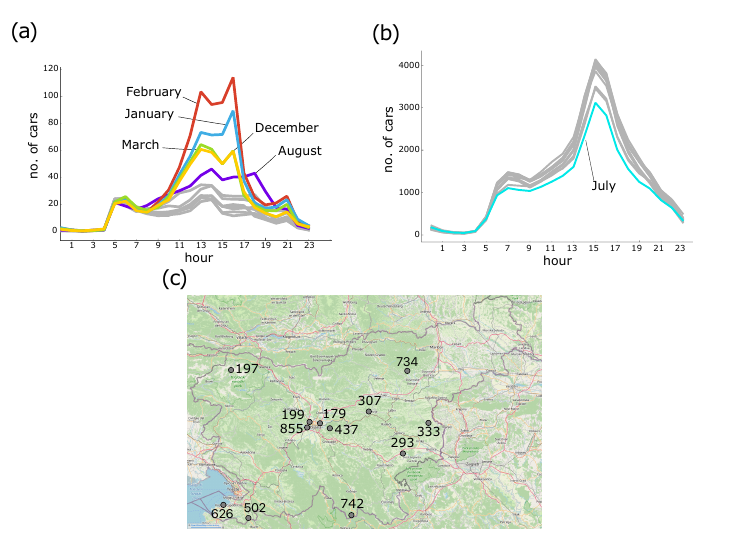}
\caption{(\textbf{a}) Counter 734 connecting \emph{Rogla} with \emph{Zreče}. In February, the two peaks represent the returnees with the half-day skiing ticket at 1 pm and those with the full-day ticket at 4 pm. The plot shows data in one direction. The average February traffic for each direction separately is shown in Figure~\ref{fig:734-feb}d. (\textbf{b}) Afternoon rush hour for counter 179. July is the month with the lowest frequency of traffic for this counter. (\textbf{c}) Map of counters that are referenced in the paper.}
\label{fig:rogla-feb} \label{fig:179-rush-hour} \label{appendix:map-of-counters} \end{figure}

The interestingness is best seen from the line plot, where each line represents the average traffic for a given period, and the x-axis indicates the hour of the day (Figure~\ref{fig:rogla-feb}a). The interestingness scores correspond to the visual information for each image. The plot would be considered interesting: if there was a considerable dispersion of lines, which is typical for seasonal trends; if there was a single outlying line, which is typical for road work; or if there was a peak at a particular hour, which indicated individual events, such as festivals, road works, \emph{etc.} We applied five statistical measures to score the counters, each corresponding to a particular type of interestingness. We used five measures of interestingness: 

\begin{itemize}
\item \emph{\textbf{A: absolute deviation from the baseline}}. Absolute deviation from the baseline takes the lowest four values as the baseline commuter traffic, then computes the difference between the individual traffic profile and the baseline.

\item \emph{\textbf{B: relative deviation from the baseline}}. The relative deviation is similar, but it looks at the ratio of difference. 

\item \emph{\textbf{C: coefficient of variation}}. The coefficient of variation~\cite{everitt2002cambridge} is a standardized measure of the dispersion of the frequency distribution. In our case, it measures how far apart from each other the traffic profiles are. Such a measure would capture counters with high variability across the data. 

\item \emph{\textbf{D: a total difference from the baseline}}. The difference from the baseline looks at each hour, computes the average traffic, and then sums the differences of the profiles to the baseline. The final score is the sum of these differences, so such a score would find counters with several high spikes. 

\item \emph{\textbf{E: adjusted z-score}}. Finally, the adjusted z-score normalizes the data to have zero mean and a variance of one. However, we replaced the mean with the baseline in our case, thus standardizing the data to the assumed everyday traffic. Z-score finds counters with singular interesting deviations.
\end{itemize}

Finding interesting traffic profiles is not just an exercise in data handling. Deviant profiles show interesting relations between the drivers and their destinations. In his account of the social life of a road between Albania and Greece, Dalakoglou~\citeyear{dalakoglou2010road} shows how roads embody both the material aspect and the socio-cultural transformation these technological objects bring to the people. However, the author's choice of the ethnographic site required previous knowledge of the importance of this particular section of the road. Measures of interestingness mitigate the need for insider knowledge by providing a way to detect potentially relevant sites objectively and exhaustively.

\section*{Detecting traffic patterns with simple methods}

We used the above scores to find interesting patterns in the road traffic data. For each method, we sorted the counters by the scores and analyzed the ten highest-ranked unique counters. We generated visualizations using Orange~\cite{JMLR:demsar13a}. Upon finding interesting counters, we did an extensive online archive search to explain why such patterns occur. Patterns began to fall into several categories, creating an image of trends and particularities of the Slovenian road traffic system.

\subsection*{High frequency counters}

Our main goal was to find interesting traffic counters, where interesting means a high traffic volume that can be seen as a total increase per month or a spike at a particular hour of the day. The first task was to find high-volume traffic counters by sorting the traffic counts for each direction and observing the highest counters. Unsurprisingly, these were the counters on the ring road around the capital that detect commuters (counters number 179, 199, 855, see Figure~\ref{appendix:map-of-counters}c). Their monthly averages also displayed a very predictable and strict weekday schedule, with incoming spikes around 7 am and outgoing spikes around 3 pm, which designate rush hours. In contrast, weekends showed a greater dispersion of the daily spike and greater variability for different hours of the day. The contrast between the weekend values of counters 179 and 855 is quite stark, with July being the lowest month for 179 and one of the highest for 855 (Figure~\ref{fig:179-rush-hour}b). That is because counter 179 is on a road leading outside of the capital and the city centre has a low traffic volume in summer. Conversely, counter 855 is located at a major motorway connecting Austria with Croatia and Italy, hence the increase in July and August.

High-frequency counters show popular transit areas or final destinations, reflecting general trends in the national traffic flow. Highways are transport lifelines, as most commuter traffic passes at least a section of it. The capital, for example, registers around 120,000 commuters daily~\cite{bajuk2017kultura}, which amounts to 6\% of the total population of the country. The initial infrastructure built to facilitate connections between regional capitals, such as Zagreb-Graz and Trieste-Budapest, serves many commuters who come to work in the capital. Due to the dependence of the traffic on highways and the country's position at the top of the Adriatic, the strain on the traffic is particularly dire in the summer. Local traffic is then forced to share the roads with neighbouring tourists.  

\subsection*{Seasonal increase and decrease}

\begin{figure}[ht]
\centering
\includegraphics[width=0.96\linewidth]{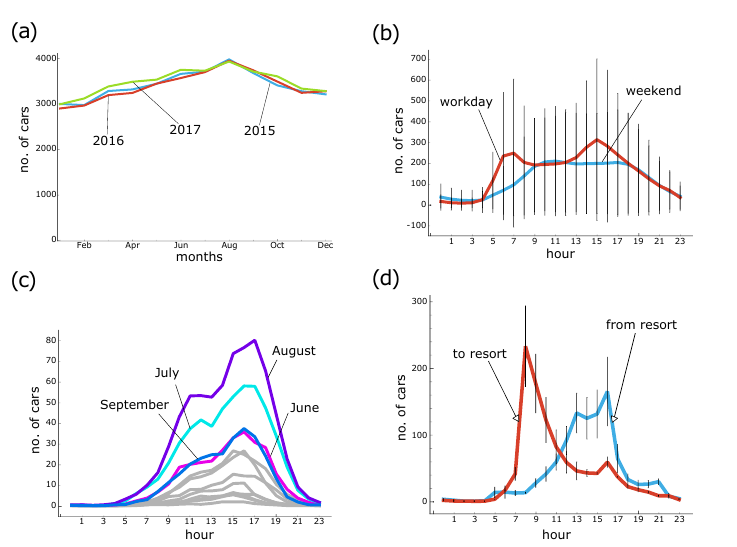}
\caption{(\textbf{a}) Average monthly traffic by year. Traffic patterns are consistent across the years, with a uniform peak in the summer. (\textbf{b}) Average car traffic per hour by the type of day, one line representing workdays and the other weekends. Vertical bars show standard deviation. (\textbf{c}) Counter 197 on the route between \emph{Vršič} pass and the \emph{Trenta} valley. The road is closed in the winter due to snowfall, but it is popular with summer visitors. (\textbf{d}) Average daily traffic for February for counter 734. The x-axis shows the hours of the day. One line shows traffic to the skiing resort, with a peak at around 8 am, while the other line shows traffic from the resort, with two peaks at 1 pm and 4 pm. Vertical bars show standard deviation.}
\label{fig:average-year-traffic} \label{fig:work-vs-weekend} \label{fig:197counter} \label{fig:734-feb} \end{figure}

Two main motorways crisscross Slovenia in an X pattern, connecting Hungary with Italy and Austria with Croatia. The intersection is in Ljubljana, with its ring road also serving daily migrants coming to work in the capital. As a transit country, Slovenia experiences high traffic in the summer, mainly in July and August (Figure~\ref{fig:average-year-traffic}a), as its northern neighbours take the journey south to Croatia.

The morning rush hour lasts from 6 am to 8:30 am, while the afternoon one starts at 2 pm and ends around 4 pm (Figure~\ref{fig:work-vs-weekend}b). Rush hour is particularly evident on Ljubljana's ring road, with high incoming traffic in the morning and a high outgoing one in the afternoon. Analyzing rush hours uncovered typical commuting patterns. The workday patterns show the country's economic hubs and when most of the workforce goes to work. Slovenia is still fairly centralized, with Ljubljana registering the highest proportion of workday traffic, with small increases in \emph{Kranj}, \emph{Koper}, and \emph{Novo mesto}. \emph{Maribor} is different, as many inhabitants work in neighbouring Austria. The rush hours are conservative, with the commute starting and finishing early in the day~\cite{anttila2015workingtime}.

However, we aimed to go beyond typical traffic flows and discover regions and locations where traffic is different from usual. We were interested in seasonal increases, short spikes, and outlying curves. We wanted to map the landscape of traffic flows quantitatively and qualitatively, and elicit patterns of behaviour that exhibit collective preferences and habits.
 
For seasonal increases, we first grouped the data into seasons. March, April, May represent spring, June, July, August represent summer, September, October, November represent autumn, and December, January, February represent winter. The final score was the difference to the mean hourly traffic. In other words, if a counter reported 30\% of annual traffic in the winter, but the overall winter traffic was 26\%, the score would be 4\% (0.04). Finally, we ranked the counters according to their interestingness as estimated by the difference to the mean and took the top 10 results.

In north-west is counter 197, counting traffic on the road from \emph{Vršič} pass to the \emph{Trenta} valley. This counter has a particularly low baseline, with very little traffic recorded over the winter months (Figure~\ref{fig:197counter}c). The drop is expected, as \emph{Vršič} is Slovenia's highest mountain pass, which is closed in the winter due to heavy snowfall. Considering it is not a vital route connecting villages or cities for the locals, it perfectly reflects the tourists' behaviour. The main peaks are around 12 am and 5 pm, showing that tourists are not early birds and prefer to finish their trip before nightfall. The traffic curve shows incoming traffic from \emph{Vršič} to \emph{Trenta} in the morning and returning traffic in the afternoon. While we cannot claim with certainty that these are the same cars, most people likely decide not to stay in the beautiful but quite remote \emph{Trenta} with fewer capacities. They instead return to \emph{Kranjska Gora}, the local tourist hub.

Counter 197 was consistently ranked at the top for the most interestingness measures, namely seasonal deviations and C and D scores. Different scores reveal different information. Score D, for example, finds counters with high relative deviation from the mean, where the score is the sum of absolute differences for each hour. This measure considers both positive and negative deviations by summing the absolute values. Moreover, each deviation from the mean adds to the score, ranking highly the counters with several deviant months. One such counter is 734, which connects the ski resort \emph{Rogla} with the spa town \emph{Zreče} (Figure~\ref{fig:734-feb}d). The traffic increase is the highest in February and January, with significant deviations in March and December as well. Both the deviations and the direction of the traffic show that \emph{Rogla} is predominantly a winter destination. The traffic peaks at 8 am going towards the ski resort, and it peaks at 1 pm and 4 pm going away from the resort. The two peaks in the afternoon correspond to two types of ski passes (half-day and full-day), each expiring exactly at the time of the traffic peak. The reason February is the most popular month is not only due to the amount of snow but also due to the winter school holidays, which occur at that time. While this destination is a typical winter destination for domestic tourists, it has recently expanded the summer offer, evident from the increase in traffic flow in August.

\begin{figure}
\centering
\includegraphics[width=0.96\linewidth]{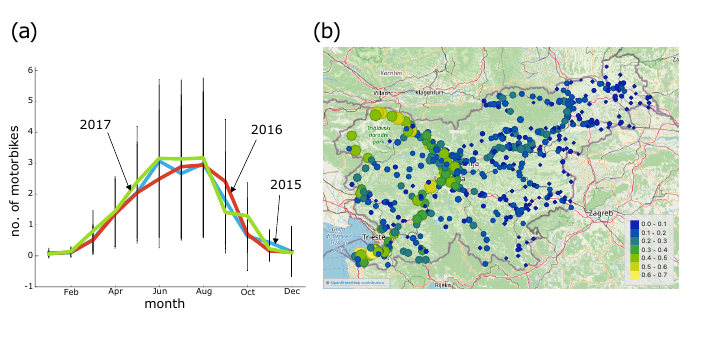}
\caption{(\textbf{a}) Average motor traffic per month. Vertical bars show standard deviation. (\textbf{b}) Relative motorbike traffic. Both the color and the size of the point correspond to the percentage of yearly traffic in September. There is high traffic in the western part of the country.}
\label{fig:avg-motor} \label{fig:motor-map} \end{figure}

Seasonal patterns are distinctive for two other entities, motorcycle riders as a type of traffic and border crossings as a type of counter. Motorcycle riders begin their season in March when there is a first significant increase in motorcycle traffic. The trend increases until June and falls slightly in July, probably because few people travel to the seaside by motorbike. The season reaches its peak in August (Figure~\ref{fig:avg-motor}a and \ref{fig:motor-map}b), then falls towards the end of the year. We can assume motorcycle riding is a seasonal leisure activity, with a slight hiatus during the summer holidays. As for the border crossings, they are, compared to other counters, at their highest in July and August. The increase starts in June at Slovenia's north-eastern (Hungary) and south-western (Italy) borders, with a similar pattern occurring in September.

Looking at seasonal peaks, we establish that tourists are unlikely to be an inconvenience on regional roads. They travel outside the main rush hour and frequent different destinations. The western part of the country is popular for tourists, especially motorcyclists, with the eastern part slowly gaining ground (particularly with tailored seasonal offers and investments in infrastructure). Local tourism is still prominent on holidays but tied to traditional destinations (\emph{i.e.}, skiing resorts in winter and Croatia in summer).

\subsection*{Weekly patterns}

\begin{figure}[ht]
\centering
\includegraphics[width=0.96\linewidth]{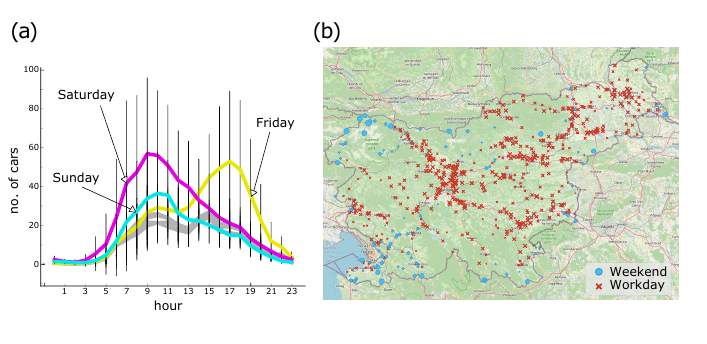}
\caption{(\textbf{a}) Daily traffic profiles for counter 742 at the southern Slovenian border. (\textbf{b}) The color and shape of the point correspond to the type of counter (blue circle for weekends and red cross for workdays), while the size corresponds to the percentage of traffic for the specified part of the week (min. 50\%, max. 70\%).}
\label{fig:742-dayofweek} \label{fig:dayofweek-map} \end{figure}

After considering seasonal fluctuations, we also had a look at weekly patterns. Instead of grouping counters by month or season, we grouped them by the day of the week. We were interested in how counters differ by daily averages, so we computed C (coefficient of variation) and D (difference from the baseline) scores for daily deviations. The top-ranked counters were two border crossings, one in \emph{Sočerga} (ID 502) in the southwest and the other connecting \emph{Fara} with \emph{Petrina} (ID 742) in the south (Figure~\ref{fig:742-dayofweek}a). These two border crossings had a strikingly similar pattern – people leaving the country on a Friday evening or Saturday morning and returning on a Sunday evening. According to unofficial data, Slovenians have around 110,000 properties in Croatia\endnote{\href{https://www.dnevnik.si/1042657187}{Slovenians' Real Estate in Croatia}}, which explains the weekend trips across the border. When observing the raw time series for the mentioned counters, we noticed that the traffic peak slowly shifts from Friday to Saturday towards the height of the summer and then back to Friday in September. The peak is likely foreign traffic since tourists from the north start their journey on a Friday and arrive at the Slovenian-Croatian border on Saturday morning. Friday traffic is predominantly due to Slovenian tourists.

Since traffic patterns seem to be highly related to the day of the week, we plotted a map (Figure~\ref{fig:dayofweek-map}b) of counters, where we coloured each counter with the part of the week, when the counter records the highest proportion of traffic. In other words, if the counter registers more traffic during the weekend than during the week, say in a 60:40 ratio, the counter is tagged as a ``weekend counter,'' and its size is 0.6. The map nicely shows typical weekend destinations with higher traffic during the weekend than during the week. The western part of the country seems popular for weekend trips, while the capital gets a lot of commuter traffic.

\subsection*{Changes in traffic infrastructure}

Most of the highest-ranked counters have increased traffic in the summer. These counters are popular tourist spots in the mountains and near the sea. Score C highly ranked one interesting counter, which had a spike in the spring. Counter 626 is located at the Slovenian coast, specifically on the scenic route between \emph{Koper} and \emph{Izola}. The route was built in 1837 and is locally known as \emph{Riva lunga} (Long coast). In March 2017, it was closed for traffic, but it was among the busiest ones before that. Considering the route's popular location, it was unclear why there was so little traffic in the summer months. However, on 5 June 2015, the tunnel \emph{Markovec} in the hinterland was finally opened\endnote{\href{https://www.rtvslo.si/slovenija/foto-gradnja-predora-markovec-je-trajala-predolgo/366795}{Building the \emph{Markovec} tunnel}}, enabling the traffic to bypass the coastal road. Longitudinally, this shows significant infrastructural changes relevant both for locals and tourists.
 
The scores detected anomalies elsewhere. Counter 307 between \emph{Zagorje} and \emph{Bevško} showed a strange increase in traffic in March. The location is not a popular tourist spot, nor was there any particular event happening to draw the crowds. The actual reason for the increase was the closure of the main road to \emph{Bevško} in March 2016 due to a landslide. The road was under reconstruction for the entire month\endnote{\href{https://www.dnevnik.si/1042732928}{\emph{Bevško} road closure}}. While this increase is not significant in the long term, it is still interesting to see how such contingencies change the landscape of traffic flows.
 
It is not just the longer traffic shifts that we were able to detect. Even traffic accidents sometimes cause enough disturbance in the typical traffic to be highly ranked for anomalies and deviations. In normal circumstances, we would prefer to discount the days with traffic accidents. Fortunately, the Slovenian Police curate a complementary data set on traffic accidents\endnote{\href{https://podatki.gov.si/dataset/mnzpprometne-nesrece-od-leta-2009-dalje}{Traffic accidents data by the Slovenian Police}}, which can be usefully linked to the traffic data to mark the accidents. Open data are not only interesting to explore on their own. They can be further enhanced with additional data sets to create heterogeneous, rich data of public life.

Changes in traffic infrastructure are indicators of economic change, while commuter flows mark the level of metropolization of the city. As new roads are built, new economic centres emerge, and old ones are abandoned, contributing to the region's decentralization~\cite{zdanowska2015metropolisation}. It marks the shift at an infrastructural and social level, with changing communication and power relations. The example of \emph{Markovec} tunnel showcases the magnitude of change when new roads take over the old ones. Still, it also shows how such a transition can be beneficial for regional development. The new road now carries the burden of traffic (commuters, freight, tourists). In contrast, the local community re-appropriated the old road to become a popular strolling path and a tourist attraction in its own right. Finding this shift in the road traffic pinpoints a starting point for future anthropological research - how were the locals socially and economically affected by the construction of the \emph{Markovec} tunnel?

\subsection*{Local festivities}

Most of our exploration focused on observing deviations from the mean, either relative or absolute. Deviations are partially reflected in the coefficient of variation, which measures how dispersed the curves are relative to the mean. We also wanted to observe smaller peaks in the curves, so we decided to compute an adjusted z-score, where instead of the deviation from the mean we observed the deviation from the baseline. Z-score~\cite{everitt2002cambridge} is a type of normalization, which puts the data on the same scale. In this way, we bypass the most frequented counters and observe local particularities. We decided to compute the z-score for each hour of each month as the deviation from a baseline. Based on initial observations, we defined the baseline commuter traffic as the mean of the lowest four values that represent the regular daily traffic. We considered only the baseline to observe interesting deviations, not the actual mean. Finally, we ranked the counters by the highest z-score per month. Here we describe three counters with high z-scores.

\subsubsection*{Chestnut festival}
 
The highest-ranked counter was 437 between \emph{Zadvor} and \emph{Šmartno pri Litiji}. The latter is a small town, while the former is a suburban section of Ljubljana. The road goes through a quaint little valley full of farms and houses. It is far from a major traffic route, so why the observed increase in traffic? Furthermore, is there a specific time in which it occurred?
 
\begin{figure}[ht]
\centering
\includegraphics[width=0.96\linewidth]{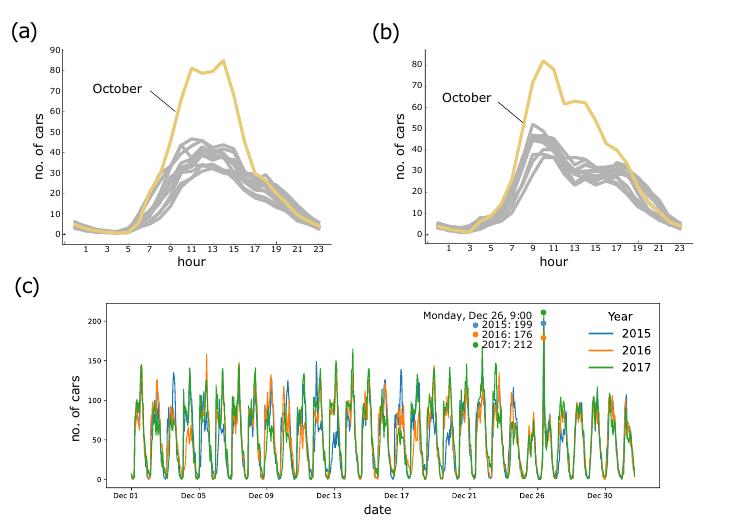}
\caption{(a) A significant increase in traffic is evident in October when \emph{Šmartno pri Litiji} with counter ID 437 holds the annual Chestnut festival. (b) A significant increase in traffic is evident in October for counter ID 333, when the Festival of \emph{Kozjansko} apple takes place in the \emph{Kozjansko} Regional Park. (c) Saint Stephen is a local, one-day event with a tiny increase in traffic at counter 293. Hence it was impossible to see the reason for the high score from the monthly averages. A daily line chart for December of all three years revealed the reason for the increase - 26 December, when \emph{Dolenja Stara vas} hold the traditional horse blessings. The image shows car traffic for December of 2015, 2016, and 2017 for one direction.}
\label{fig:437-chestnut} \label{fig:333-kozjansko} \label{fig:293-stephen} \end{figure}

Looking at the graph (Figure~\ref{fig:437-chestnut}a), it seems like October is responsible for a high z-score. Moreover, the increase happens on the weekend and follows a typical incoming-outgoing pattern. People seem to drive from the direction of Ljubljana towards \emph{Šmartno pri Litiji} between 11 am and 2 pm, while they all seem to be returning around 5 pm. The counter does not display such behaviour for other months or even for working days in October. The reason for this increase is the annual festival of chestnut\endnote{\href{https://www.dnevnik.si/1042787995}{Chestnut Festival}}, held in the hill-top village of \emph{Janče}. Inhabitants of the capital seem drawn to the nearby celebration of this autumn delicacy, showing the importance of regional festivities for domestic tourism, particularly outside of the summer season.

\subsubsection*{The festival of Kozjansko apple}
 
Among the top results is the counter 333 leading from \emph{Šonovo} to \emph{Podsreda} (Figure~\ref{fig:333-kozjansko}b). While the authors of this paper knew about \emph{Podsreda} due to its beautiful castle, the village \emph{Šonovo} was quite unknown. It turned out that the road is located in the \emph{Kozjansko} Regional Park, a protected natural area in the South-East of Slovenia. The park hosts a festival of the \emph{Kozjansko} apple\endnote{\href{https://kozjanski-park.si/?event=20-praznik-kozjanskega-jabolka}{Festival of the \emph{Kozjansko} apple}} in October, hence the increase. The incoming-outgoing pattern corresponds to the time of the festival.

\subsubsection*{Celebration of Saint Stephen}

Finally, we were able to detect a small local holiday, the celebration of Saint Stephen (\emph{Štefanovo}) in \emph{Dolenja Stara vas}\endnote{\href{https://www.rtvslo.si/zabava/novice/na-stefanovo-tradicionalni-blagoslovi-konj/510173}{Saint Stephen's Horse Blessings}} at counter 293 (Figure~\ref{fig:293-stephen}c). The holiday takes place every 26\textsuperscript{th} of December, with the traditional blessings of horses and a parade. The festival is 161 years old and registers increasing interest in observing this local tradition.

The final result was, however, not among the top-ranked. The increase is relatively small and short and will always be overtaken by larger relative increases, especially summer ones. The score should be adjusted to detect small and narrow peaks by considering a broader temporal window and increased variance per hour. That said, z-scores, while imperfect in their current implementation, nicely identify local peculiarities and interest points for domestic tourism. These increases were detected outside the main tourist season, especially in October, the main month for festivities related to crops and produce. 

The findings show the importance of domestic flows for smaller, lesser-known points of interest, which cannot compete with traditional tourist hot spots in the main season. In their modern interpretation, local festivities are a part of the cultural heritage, which benefits the locals socially and economically. As Poljak Istenič~\cite{poljak2013druzbeno} argues, the revitalization of these festivals is a good practice for the sustainable development of the peripheral and rural communities. Future research could focus on the role of off-season festivities for local communities, with a case study of one of the proposed counters. The computational analysis thus pinpointed a relevant research question and indicated potential field sites. With additional data sets on the longitudinal economic performance of the proposed locations, their demographic composition, or car-level tracking records, one could also describe changes in time and the effects of these festivals on the local community. However, the quantitative analysis did not answer why tourists visit these particular festivals, who visit them, and their social and cultural role in local life.

\section*{Detecting traffic patterns with clustering}

In the second part of the analysis, we used data clustering to discover typical traffic patterns in the country. We took the data for May, as this is the month before the main tourist season, but still popular with the local tourists. The data reported the percentage of traffic for each counter per hour, with a distinction between weekends and weekdays and the direction of traffic.

We used two clustering methods, k-means with Euclidean distance and hierarchical clustering with the Spearman correlation coefficient. For k-means, we estimated the optimal number of clusters with silhouette scoring~\cite{rousseeuw1987silhouettes}, and the result was two clusters (silhouette score=0.76). For hierarchical clustering, we used Ward linkage and estimated the number of clusters from the dendrogram - the inter-cluster distances started increasing between four and six clusters. Finally, we grouped the counters by cluster labels to compute the average cluster profile and plotted the results in a line plot. The choice of six clusters was appropriate, as other cutoffs resulted in overly similar cluster profiles.

\begin{figure}[ht]
\centering
\includegraphics[width=0.96\linewidth]{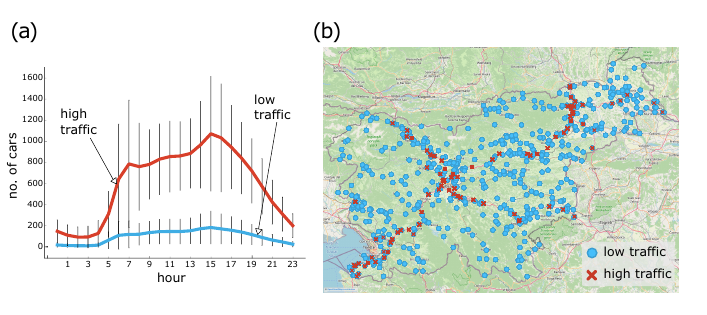}
\caption{(\textbf{a}) Average traffic profiles for the two clusters found with k-Means. They correspond to high and low traffic counters. Vertical bars show standard deviation. (\textbf{b}) Map of counters as clustered by k-Means. Blue circles represent low traffic counters, while red crosses represent high traffic counters.}
\label{fig:k-means-clustering} \label{appendix:k-means-map} \end{figure}

K-means found two distinct clusters, which correspond to the high traffic and low traffic counters (Figure~\ref{fig:k-means-clustering}a). The result is unsurprising since the magnitude of the measured phenomena heavily influences Euclidean distance calculation. The high traffic counters are located on the so-called highway cross of the country, the four main motorways joining in the capital (Figure~\ref{appendix:k-means-map}b).

\begin{figure*}[ht]
\centering
\includegraphics[width=0.7\textwidth]{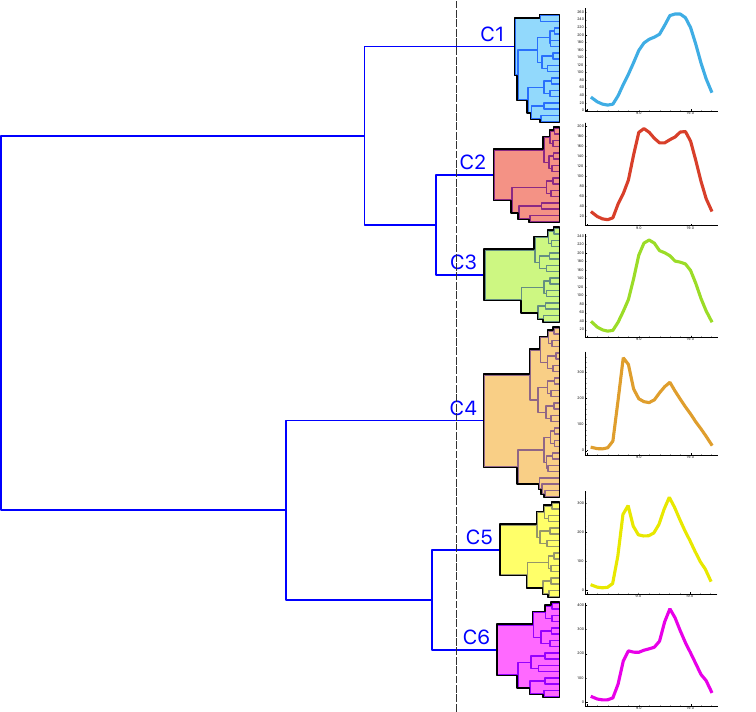}
\caption{A dendrogram of traffic profiles for May. Clustering is performed with Spearman correlation coefficient and Ward linkage. The dendrogram is pruned at split depth 7 for a concise view. An average line plot is shown next to each cluster.}
\label{fig:hierclust-dendrogram}
\end{figure*}

Hierarchical clustering with the Spearman correlation coefficient gave us a different image. We used the Spearman correlation coefficient, which measures how similar the shapes of the curves of traffic data are~\cite{ye2015time}. The sensible cutoff was somewhere between two and six clusters. The six clusters corresponded to the workday and weekend profiles and the morning and afternoon rush hour (Figure~\ref{fig:hierclust-dendrogram}). Setting the cutoff at two clusters corresponds to the workday and weekend clusters, meaning there is a significant difference between the two periods. With three clusters, the workday cluster is further split into two groups: the morning and the afternoon rush hour. With four clusters, the weekend curves are further split into morning and evening traffic.

The first cluster contains predominantly weekend profiles (76.16\% weekend, 23.84\% workday), with a slow increase in traffic between 4 pm and 5 pm. The second cluster is almost exclusively a weekend cluster (98.92\%) and has higher traffic in the early morning. The third cluster is also a weekend cluster (95.79\%), but in this case, the traffic is higher in the wee hours, namely, at midnight and one and two in the morning. The fourth cluster contains almost exclusively workday profiles (99.78\%), including the morning rush hour, with higher values for 5 am, 6 am, and 7 am. The fifth cluster is exclusively a workday cluster (100\%), with a morning and afternoon spike. The sixth cluster is also a workday cluster (93.76\%), with higher traffic between 2 pm and 4 pm - the afternoon rush hour.

The results are very similar for all other months. Winter months have less distinct weekend traffic patterns and more distinct patterns during the week. Interestingly, December has a high afternoon peak, which contains about half of the workday and half of the weekend counters, reflecting the afternoon errands and festivities typical for this month. The weekend afternoon spike shifts to later hours in the summer, while the morning spike shifts to earlier hours and becomes a separate cluster. The morning cluster predominantly consists of the tourist traffic that transits Slovenia. Finally, the average weekend traffic in July and August overtakes the average workday traffic.

When selecting three clusters, regardless of the month chosen, the most typical traffic patterns were the workday \emph{vs.} weekend curves and the afternoon rush hours. The workday patterns are relatively typical, with spikes at 6 am and 3 pm. The weekend curve starts the climb later, at around 10 am, continues throughout the day, and drops at around 4 pm. The distinctive weekend counters are at the border and more frequently in the western part of the country. Slovenians seem to travel across the border for the weekend frequently. If they stay in the country, the \emph{Primorska} region appears to be the most popular (the same is true for tourists coming into Slovenia).

\section*{Results of quantitative analysis}

In the paper, we show how to use open public data for anthropological research. Specifically, we show how computational approaches such as data mining and machine learning can be used for finding interesting patterns of human behaviour, which are then qualitatively interpreted and explained. The focus is not on data mining and machine learning, which have already proven valuable in anthropological research~\cite{blok2014complementary,krieg2017anthropology,pretnar2018power,pretnar2019data}, but on the potential of quantitative analysis in anthropology and the rich information hidden in public data sources.

Simple statistical methods are great for initial data exploration and uncovering interesting patterns. Certain interestingness scores reveal general traffic trends (scores A and B). Others detect local deviations (C, D, and E scores), such as road work, seasonal festivities, and popular tourist spots. The data is recorded at hourly intervals, so it is easy to observe trends for a single traffic counter or group in different periods. The ability to traverse between different levels of granularity is one of the key benefits of using sensor data for anthropological research. Identifying outlying patterns enables preliminary analysis of the phenomenon (in this case, traffic) and pinpointing potentially relevant field sites.

For a comparative analysis of traffic patterns, we used clustering. Cluster analysis shows that traffic patterns differ primarily by the day of the week and the time of the day. Patterns are not location-specific - latitude and longitude of the traffic counter do not play a role in distinguishing between clusters. In other words, there is no region where a specific pattern would be predominant. The only exceptions are the western part of the country for motor traffic and the border counters with increases at the end of the week. This approach showed the general structure of the road network, which is the basis for understanding national traffic flow, not only materially and logistically but also socially. Clustering reveals patterns of frequent behaviour, categorizes human practices, and enables subsequent comparative analysis.

Analysis of road traffic counters revealed a significant distinction between workday and weekend traffic patterns. Most people still seem to work predominantly from about 7 am to 3 pm. For the weekend, they go for a trip at around 10 am, then stop for lunch at around 1 pm, and return home by 5 pm. Many people travel to Croatia for the weekend, leaving on Friday evening or Saturday morning and returning on Sunday evening. Slovenia is still a transit country with an increase in traffic on the highways in summer. That said, summer tourism is well established in the western part of the country. Some winter destinations are being transformed to year-round destinations, as is the case with the \emph{Rogla} ski resort. Off-season activities are also popular with the locals. In October, people attend different autumnal events that celebrate local produce and traditions, while in December, there is an emphasis on religious festivities and pre-New Year's Eve celebrations.

The present study has certain limitations. The data is not related to individual activity but to the aggregation of local mobility patterns, making it impossible to observe individual behaviour patterns. It is also impossible to completely discount extraneous factors affecting counter data (such as road closures or local points of interest). These factors were already partially handled by aggregating over a longer time (3 years), but the data could be expanded to an even longer time frame. In some specific cases, however, over-zealous aggregations discount small local events such as the celebration of St. Stephen. Aggregations thus have to be handled case-by-case.

With the identification of general, specific, and local traffic properties, we have shown a glimpse into public life as seen from the road. The anthropological study of traffic should highlight universals in behaviour as well as cross-cultural differences in intention, interpretation, interaction, and management of risk~\cite{rosin2003vishnu}. The present study, experimental in its nature, was able to pinpoint certain universals if we consider universals as general traffic patterns found throughout the world~\cite{pasaoglu2012driving,wolday2019workplace}. The study also revealed intentions and interactions by studying deviations from the general patterns. 

Traffic data is highly anonymized and aggregated at 15-minute intervals. Without additional data, such as the country of origin or individual car movement, one can only get a general overview of national traffic. Aside from archival sources, it is almost impossible to relate these data to lived experiences~\cite{wilmott2016small}, making them ethnographically sparse. That said, we were able to outline the general traffic of Slovenia with some interesting specifics that were investigated and explained with archive resources. Such experimental analyses are, in our opinion, essential for pushing the limits of a single field (\emph{i.e.}, anthropology) and for finding interesting interdisciplinary intersections and cooperation.

\section*{Quantitative anthropology as the future?}

With the broad availability of open data, it is easier than ever to analyze vast amounts of data on human practices computationally. Quantitative data reveals the structure of the phenomena, temporal changes, and interesting outliers, all highly relevant for describing a phenomenon or a community in detail. Nevertheless, computational analysis insufficiently addresses some of the key questions of anthropology - why and how? It does not offer detail-rich descriptions and context, and it does not and cannot relate correlations to the cause. That said, computational anthropological research is a great starting point and can provide a relevant glimpse into the life of a community.

Quantitative analyses of mobility data provide information on general mobility patterns and seasonal trends. They also reveal interesting outliers, which can serve as a starting point for forming research questions. Detecting interesting locations can also help the researcher narrow down potential research sites. Finally, ethnographic observations of mobility can be supplemented with findings from quantitative studies. Interdisciplinary approaches often provide a richer picture of a phenomenon than a single method~\cite{greene2010dialectics}.

Moreover, open data are a great starting point for quantitative research. Such data can hold interesting information on people's behaviours, practices, and habits. One of the most attractive opportunities in the age of digital data is finding quality in quantity, or in other words, reflections of social structures, cultural patterns, norms, and values in numerical data. Open data is essentially archival. We argue that it benefits the research by studying the community in a broader context, detecting rare events, observing temporal cycles, and mitigating biases.

Analysis of traffic data revealed several strengths and weaknesses of quantitative approaches for anthropological analysis. Computational techniques help find general patterns in large data sets and interesting deviations~\cite{krieg2017anthropology,pretnar2019data}. Large data sets can be analyzed on several levels. One level is the overview, taking the entire data set and observing its properties. Such analysis reveals typical behaviours, relations, and hierarchies of the population. In the case of traffic data, the overview would entail nationwide traffic patterns, seasonal fluctuations, and clustering of traffic profiles. Another level is the midway analysis, with the observation of data segments, sub-populations, regions, \emph{etc.} An example from the traffic data analysis would be comparing different parts of the country, looking at the inbound-outbound traffic of a city, or observing specific periods. The most detailed level is the granular analysis of individual data points or categories. We observed specific counters longitudinally and comparatively (in terms of temporal or spatial differences). These different levels of analysis are not specific to quantitative data. However, they are much more pronounced in this case since traversing between levels must be explicitly encoded in the data analysis procedure. Finally, quantitative analyses are great at answering ``where,'' ``what,'' ``when,'' and ``who.''

However, determining ``how'' and, most importantly, ``why'' with quantitative data is more challenging. Traffic data, for example, is not very rich in detail. It is aggregated at a 15-minute interval, bound to a specific counter, and contains only indirect data on human habits. To answer ``why'' a particular behaviour occurs, we had to resort to archive data and newspaper clips. The ``how'' is partially reflected in the routes taken and the time driving occurs, but again, the aggregated data cannot accurately reflect individual variability.

Consequently, we should substantiate open data with qualitative data to achieve a valid ethnographic account and provide a detail-rich explanation of traffic practices. However, the quantitative analysis does pinpoint broad behaviours that serve as a starting point for further research. Bajuk Senčar~\citeyear{bajuk2017kultura} argues that traffic infrastructure can be understood as a system and a process. Analysis of traffic data describes the system, which serves as a base for understanding the process. Furthermore, open data sets are a great resource, as they are readily available, well-structured, and accompanied by metadata. However, the process should be explained ethnographically by connecting the materiality of the infrastructure with social actors, that is, the people behind the process.

Anthropological research using open public data sets is not meant to replace ethnographic fieldwork but should be considered a complementary venue with numerous possibilities. It is mainly appropriate for preliminary analysis, as shown by identifying interesting traffic counters. Moreover, it provides insight into the structure of the phenomenon or a system. It shows when an event occurs, where, and how. In other words, it describes the phenomenon in terms of its properties, but, as explained above, it cannot provide the context and the reason why something happens. Nevertheless, as computational analyses are time and resource-efficient and reveal relevant social and cultural aspects, they are a valuable research tool.
 
\section*{Availability of data and materials}
Jupyter notebook scripts, Orange workflows, and data that support the findings of this study are available in figshare with identifier  \url{https://doi.org/10.6084/m9.figshare.13259093}.

\section*{Competing interests}
The authors declare that they have no competing interests.

\section*{Funding}
This work was supported by the Slovenian Research Agency [P2-0209] and co-funded by the Slovenian Ministry of Education, Science and Sport, and the European Regional Development Fund [Tourism 4.0 - enriched tourist experience (OP20.03536)]). The funding bodies did not influence the design of the study, the collection, analysis, and interpretation of data, or the writing of the manuscript.

\section*{Author's contributions}
APŽ and TC designed the study. APŽ and TC acquired the data. APŽ wrote most of the data analysis scripts. APŽ, TH and TC analyzed the data. APŽ wrote the first draft. APŽ, TH and TC wrote and reviewed the manuscript. All authors read and approved the final manuscript.

\section*{Acknowledgements}
The authors thank Urška Starc Peceny and Samo Eržen for feedback on the research questions and design addressed in this paper. 
\bibliographystyle{unsrt}

\bibliography{main}

\end{document}